\documentclass[prd,superscriptaddress,showpacs, twocolumn, nofootinbib]{revtex4-1}

\usepackage{amsmath, amssymb,bbm}
\usepackage{braket}						
\usepackage{graphicx} 					
\usepackage{hyperref}					
\usepackage[usenames, dvipsnames]{color}	
\usepackage[usenames,dvipsnames]{xcolor}
\usepackage[caption=false]{subfig}
\usepackage{color}
\usepackage[normalem]{ulem}

\newcommand{\abs}[1]{\left| #1 \right|} 						
\DeclareMathOperator{\tr}{tr} 								
\newcommand{\diff}[1]{\ensuremath{\operatorname{d}\!{#1}\ }}		
\renewcommand{\Im}{\mathop{\rm Im}}						
\DeclareMathOperator\sgn{sgn}							
\DeclareMathOperator\erf{erf}								
\DeclareMathOperator\erfc{erfc}							
\DeclareMathOperator\PV{PV}								
\newcommand{\dif}[1]{\operatorname{d}\!{#1} \,}	

\def\1{{{\mathbbm 1}}}
\def\6{{\langle}}
\def\9{{\rangle}}

\def\be{\begin{equation}}
\def\ee{\end{equation}}

\def\bes{\begin{align}}
\def\ees{\end{align}}

\def\ra{\rightarrow}

\def\cc{{\mathcal{C}}}
\def\cn{{\mathcal{N}}}
\def\co{{\mathcal{O}}}
\def\mv{{\mathrm{v}}}

\newcommand{\magenta}{\textcolor{magenta}}

\newcommand{\nn}{\nonumber}

\usepackage{soul}

\DeclareMathOperator*{\SumInt}{%
\mathchoice%
  {\ooalign{$\displaystyle\sum$\cr\hidewidth$\displaystyle\int$\hidewidth\cr}}
  {\ooalign{\raisebox{.14\height}{\scalebox{.7}{$\textstyle\sum$}}\cr\hidewidth$\textstyle\int$\hidewidth\cr}}
  {\ooalign{\raisebox{.2\height}{\scalebox{.6}{$\scriptstyle\sum$}}\cr$\scriptstyle\int$\cr}}
  {\ooalign{\raisebox{.2\height}{\scalebox{.6}{$\scriptstyle\sum$}}\cr$\scriptstyle\int$\cr}}
}

\begin{document}

\preprint{}

\title{Spacetime structure and vacuum entanglement}

\author{Eduardo Mart{\'i}n-Mart{\'i}nez}

\affiliation{Institute for Quantum Computing, University of Waterloo, Waterloo, Ontario, N2L 3G1, Canada}
\affiliation{Department of Applied Mathematics, University of Waterloo, Waterloo, Ontario, N2L 3G1, Canada}
\affiliation{Perimeter Institute for Theoretical Physics, 31 Caroline St. N., Waterloo, ON, N2L 2Y5, Canada}

\author{Alexander R. H. Smith}
\email[]{a14smith@uwaterloo.ca}
\affiliation{Department of Physics \& Astronomy, University of Waterloo, Waterloo, Ontario Canada N2L 3G1}
\affiliation{Department of Physics \& Astronomy, Macquarie University, NSW 2109, Australia }

\author{Daniel R. Terno}
\affiliation{Department of Physics \& Astronomy, Macquarie University, NSW 2109, Australia }

\date{\today}

\begin{abstract}

We study the role that both vacuum fluctuations and vacuum entanglement of a scalar field play in identifying the spacetime topology, which is not prescribed from first principles---neither in general relativity or quantum gravity. We analyze how the entanglement and observable correlations acquired between two particle detectors are sensitive to the spatial topology of spacetime. We examine the detector's time evolution to all orders in perturbation theory and then study the phenomenon of vacuum entanglement harvesting in Minkowski spacetime and two flat topologically distinct spacetimes constructed from identifications of the Minkowski space. We show that, for instance, if the spatial topology induces a preferred direction, this direction may be inferred from the dependence of correlations  between the two detectors  on their orientation. We therefore show  that vacuum fluctuations and vacuum entanglement harvesting makes it, in principle, possible to distinguish spacetimes with identical local geometry that differ only in their topology.

\end{abstract}


\maketitle

\section{Introduction}
\label{Introduction}


The field equations of general relativity are local. As such, they describe the local structure of spacetimes, while  remaining silent about the large scale structure of our Universe, including its spatial topology \cite{Hawking:1973}. Different cosmological models produce a variety of global properties, including a range of spacetime topologies. In principle, we do not expect a future theory of quantum gravity to fix the spatial topology of the Universe. For example, loop quantum gravity admits the possibility of topological changes \cite{Thiemann:2007zz}, while path integrals in spin-foam  models are summed over all possible topologies \cite{Engle:1982}.  How our actual observable Universe behaves, from the point of view of topology may be derivable from the initial conditions of the Universe \cite{Martin:2013}, but at present we believe that the underlying fundamental theories are compatible with whatever large scale structure we happen to live in.

It is then a task of observational astrophysics and cosmology to determine the actually realized scenario \cite{Ade:2013vbw, *Ade:2013xla}. Unfortunately, to date, astronomical observations of the large scale structure of the Universe are compatible with either an Euclidian spatial topology or any kind of open or closed topology where the relevant topological scales are comparable or above the Hubble radius. Even if the signals we are collecting may have originated in causally disconnected regions, our entire measurement record is overwhelmingly local in space and time.

Quantum mechanics possesses inherently non-local properties. The paramount manifestation of those properties is embodied by quantum entanglement \cite{Peres:2004, Bruss:2007, *Nielsen:2010}. The fundamental question that we address in this paper is: Can we use quantum mechanics to amplify the non-locality of our experiments and thus become more sensitive to aspects such as the topology of spacetime?

From the point of view of local observers the vacuum state of any quantum  field is entangled, and thus localized vacuum fluctuations are correlated \cite{Peres:2004}. While the Unruh effect \cite{Davies:1975,Fulling:1973,Unruh:1976} is perhaps the most spectacular manifestation of vacuum entanglement, acceleration of observers is not  required. It was demonstrated that correlations in the vacuum measured by local inertial observers are, in principle, strong enough to violate Bell-type inequalities \cite{Summers:1985, *Summers:1987fn}. Furthermore, it is indeed known that localized particle detectors can extract entanglement form the vacuum state of a quantum field, even while remaining spacelike separated, due to the intrinsic non-locality of the field ground state \cite{Valentini:1991,Reznik:2002fz,Hu:2012jr,Pozas-Kerstjens:2015}. This phenomenon has become known as `entanglement harvesting' \cite{Salton:2014jaa},  and it has already been proven to encode information about the global structure of spacetime. Specifically,   \citet{Steeg:2009} demonstrated that entanglement harvesting is sensitive to the large scale structure of spacetime.

{Continuing the study} of the dependence of vacuum entanglement of a quantum field on cosmological parameters, we consider a real massless scalar field and a pair of two-level Unruh-DeWitt detectors \cite{DeWitt:1979, Crispino:2008}  and derive their density matrix assuming only existence of the Wightman function. Using Minkowski spacetime as a benchmark, we compare the resulting entanglement between the two detectors, with the counterparts in topologically non-trivial locally flat spacetimes.


This paper is organized as follows: In Section \ref{Detectors in Minkowski Space} we review the detector model and derive the joint density matrix of two detectors in Minkowski space, initially in their ground state and at rest with respect to one another, to all orders of perturbation theory and present a consistent leading-order expression.  We then analyze the probability that a single detector undergoes a transition to the excited state, and quantify the entanglement harvested between the two detectors resulting from interacting with the field.  In Section \ref{Detectors in quotients of Minkowski space} we consider an analogous situation in two locally flat, topologically distinct, quotient spacetimes constructed from Minkowski space. Since entanglement measures are locally unobservable, we evaluate and compare, in addition to local statistics, the observable correlations between the detectors. We conclude in Section \ref{Discussion and outlook} with a discussion of the results presented and outline future directions of research.

We use the $+$$-$$-$$-$ convention for the signature of the metric and work in natural units $\hbar=c=1$.

\section{Detectors in Minkowski Space}
\label{Detectors in Minkowski Space}

In this section we analyze general properties of two Unruh-DeWitt detectors $A$ and $B$ that interact locally with a real massless scalar field for (effectively) a finite period of time. We assume that the spacetime in question allows us to separate solutions of the Klein-Gordon equation into positive- and negative-frequency modes, thus allowing  us to define the Wightman function \cite{Fulling:1973},  and that the two detectors are situated in the same coordinate patch. This is enough to obtain the structure of their joint detector density matrix $\rho_{AB}$ \big(Eq.~\eqref{rhoAB}\big), and a general form of its matrix elements and entanglement properties. Specific information about investigated spacetimes enables us to calculate this expressions explicitly. While our analysis is based on perturbation theory, we obtain results that are valid at all orders.  As a benchmark we provide the leading order expressions for two identical detectors at rest in Minkowski space.

\subsection{The detector model}
\label{The detector model}

We will model our particle detectors as two-level quantum systems that interact with the real massless scalar field $\phi$ via an Unruh-DeWitt monopole coupling \cite{DeWitt:1979}. Under this coupling model, the time-dependent interaction Hamiltonian in the interaction picture is given by
\begin{align}
H_I \left(\tau\right) = \lambda\! \left(\tau\right) \left(e^{i\Omega \tau}\sigma^+ + e^{-i\Omega\tau} \sigma^- \right)  \phi\left[x(\tau)\right], \label{InteractionHamiltonian}
\end{align}
where $\tau$ is the proper time of the detector, $\lambda\left(\tau\right)$ is a weak time-dependent {coupling parameter} which controls the strength and length of the interaction, $\Omega$ is the energy gap between the ground state $\ket{0}_d$ of the detector and its excited state $\ket{1}_d$, $\sigma^{\pm}$ are SU(2) ladder operators which act on the state of the detector, that is $\sigma^+ \ket{0}_d = \ket{1}_d$, $\sigma^- \ket{1}_d = \ket{0}_d$, ${(\sigma^{\pm})}^2=0$ and $\phi\left(x(\tau)\right)$ is the field evaluated along the trajectory of the detector. This model, while simple, captures most of the relevant features of the light-matter interaction when no angular momentum exchange is involved \cite{Martin-Martinez2013,Alvaro}.

{As we will assume the detectors are at rest with respect to one another}, it is convenient to parameterize both detectors and the field's time evolution by the common coordinate time $t$. We {express the coupling parameter as $\lambda(t) =  \epsilon_0 \epsilon(t)$, where $\epsilon_0\ll 0$ is the coupling strength and $\epsilon (t) = e^{-t^2/2\sigma^2}$ is a Gaussian switching function.}  Due to the strong suppression for $\abs{t}\gg \sigma$, we can approximate the detector being ``on'' when $\abs {t} \lesssim \sigma$ and ``off'' otherwise.
%

Prior to the interaction, we consider that the detectors are initially in their ground states $\ket{0}_A$ and $\ket{0}_B$, and the field is in the vacuum state $\ket{0}$, so that the initially joint state of the two detectors and field is given by $\ket{\Psi} = \ket{0}_A\ket{0}_B\ket{0}$. During the interaction the composite system undergoes the unitary evolution
\begin{align}
U = \hat{T}  e^{-i \int\dif t  \big[ H_A(t) +H_B(t) \big] }  , \label{jointUnitary}
\end{align}
where $\hat{T}$ denotes time ordering and the Hamiltonians $H_A$ and $H_B$ describe the field interaction with detectors $A$ and $B$, respectively.  These Hamiltonians are given by Eq.~\eqref{InteractionHamiltonian}. In principle, the detectors may have different interaction coupling strengths, switching functions, and energy gaps. Even when considering identical detectors we will use the subscripts $A$ and $B$ for the coupling parameters $\epsilon_A=\epsilon_B=\epsilon_0$,  since it allows us to easily distinguish local and non-local terms in the perturbation expansion below. The two interaction Hamiltonians commute at equal times, $[H_A(t),H_B(t)]=0$.

Some of the calculations are more conveniently performed using the Schr\"{o}dinger picture. We label the basis state of the detectors as $m_{A}, m_B \in \{0,1\}$ and the states of the field as $\mu$.  The combined system is given by
\begin{align}
\ket{\psi\left(t\right)} &= \SumInt_{m_A, m_B , \mu}  c_{m_A m_B\, \mu}(t) \nn\\
&\qquad\qquad e^{-i(E_{m_A}+E_{m_B})t}   \ket{m_Am_B} U_\phi(t)\ket{\mu},
\end{align}
where the (generalized) basis states $\ket{\mu}$ of the scalar field are freely evolved by a unitary operator $U_\phi$, and the summation and integration with appropriate measures are performed over the entire multimode Fock  space. Hence the index $\mu$ can be 0 (the vacuum state), or take a single- or multi-mode labels. The coefficients $c_{k_A k_B\, \nu}(t)$ are the solutions of
\begin{widetext}
\begin{align}
i\dot{c}_{k_A k_B\, \nu}(t) &=   \left[ \SumInt_\mu  c_{k_A+1, k_B\, \mu}(t)e^{(-1)^{k_A+1}i  \Omega_A t }  \epsilon_A(t) \Braket{ \nu|\phi\left[x_A(t)\right] | \mu}   \nn \right. \\
&\left. \qquad \qquad \qquad \qquad \qquad  +  \SumInt_{ \mu}  c_{k_A, k_B+1\, \mu}(t)e^{(-1)^{k_B+1} \Omega_B t }  \epsilon_B(t) \Braket{ \nu|\phi \left[x_B(t)\right] | \mu} \right], \label{cpattern}
\end{align}
\end{widetext}
where we use mod 2 addition for the detector indices $k_A$ and $k_B$.

{We will assume the two detectors are initially unexcited and the field to be in the vacuum state, which corresponds to the initial state $\ket{\Psi} = \ket{0}_A \ket{0}_B \ket{0}$ }. This initial state corresponds to a single non-zero coefficient $c^{(0)}_{000}=1$ {at the zeroth order in the perturbative expansion of the coefficients}. Beginning with the zeroth order, higher order corrections to the coefficients $c_{k_A k_B\, \nu}(t)$ are obtained by substituting coefficients of one lower order on the right hand side of Eq.~\eqref{cpattern}.  In what follows we take $t\to \infty$, which yields the coefficients in the late time limit when the interaction between the detectors and field has ceased.

 The joint state of the two detectors  is $\rho_{AB} = \tr_\phi [ U \ket{\Psi} \!\bra{\Psi} U^\dagger]$, where the trace is over the {field degrees of freedom}.   It follows from Eq~\eqref{cpattern} that the $n$th order coefficient $c^{(n)}_{k_Ak_B\mu}$, where $\mu$ involves $m\leq n$ particles, depends on the coefficients of the order $n-1$ with the opposite detector parity sum $k_A+k_B$ and $m\pm1$ particles. As a result, at all orders of perturbation theory the density matrix
\begin{align}
 \left(\rho_{AB}\right)_{k_A l_B,m_A n_B}:= \SumInt_{ \mu}c_{k_A l_B\mu}c_{m_A n_B \mu}^* \label{MatrixElements}
\end{align}
has the form of the so-called X-state \cite{ali:10},
\begin{align}
 \rho_{AB}=\begin{pmatrix}
r_{11} & 0 & 0 & r_{14}e^{-i\xi} \\
0 &r_{22} & r_{23}e^{-i\zeta} & 0 \\
0 & r_{23}e^{i\zeta} &r_{22} & 0\\
r_{14}e^{i\xi} & 0 &0 & r_{44},
\end{pmatrix} \label{rhoAB}
\end{align}
in the basis $\left\{ \ket{00},  \ket{01}, \ket{10}, \ket{11} \right\}$ where $\ket{ij} = \ket{i}_A \ket{j}_B$, and all the coefficients $r_{ij}$ are positive. Since {$\rho_{AB}$} is a valid density matrix, the normalization condition $\sum_i r_{ii}=1$, and the following two positivity conditions must be satisfied:
\begin{align}
r_{11}r_{44}\geq r_{14}^2, \qquad r_{22}r_{33}\geq r_{23}^2.
\end{align}

A useful form of this matrix that explicitly separates the local and nonlocal quantities is
\begin{align}
\rho_{AB}=\begin{pmatrix}
1- A-B+E & 0 & 0 & X \\
0 &B-E & C & 0 \\
0 & C^* &A-E & 0\\
X^* & 0 &0 & E
\end{pmatrix}, \label{TwoDetectors}
\end{align}
Indeed, tracing out either of the detectors in the state $\rho_{AB}$, say detector $B$, results in the state $\rho_A$ of detector $A$
\begin{align}
\rho_{A}=
\begin{pmatrix}
1- A& 0  \\
0 & A
\end{pmatrix}, \label{SingleDetector}
\end{align}
in the basis $\{ \ket{0}_A, \ket{1}_A \}$. The same result is obtained when we analyze a single detector that interacts with the field via the Hamiltonian $H_A$.
If the appropriately defined  separation between the detectors (e.g.  distance $L$ between the stationary detectors) increases, they become virtually independent and the total state approaches the direct product of the density matrices of the individual detectors,
\begin{align}
\rho_{AB}=\begin{pmatrix}
1- A& 0  \\
0 & A
\end{pmatrix} \otimes
\begin{pmatrix}
1- B& 0  \\
0 & B
\end{pmatrix}.
\end{align}
As a result, $X\to 0$, $C\to 0$, and $E\to AB$.

To simplify the exposition we consider  the case of two  identical detectors, i.e., $H_A=H_B$. Hence, in translation-invariant spacetimes we substitute $B\rightarrow A$ in Eq.~\eqref{TwoDetectors}. In the leading order of the perturbation parameters $\epsilon_A,\epsilon_B$, the quantities $A$, $X$, $C$ and $E$ are given by
\begin{align}
A &= \epsilon_A^2 \int_{-\infty}^{\infty} \diff t \!\int_{-\infty}^{\infty} \dif{t'} \epsilon \left(t\right) \epsilon \left( t' \right) \nn \\
&\qquad \qquad  e^{-i\Omega \left(t - t'\right)} W \left(x\left(t\right), x\left( t' \right)\right) + \mathcal{O}\!\left(\epsilon_0^4\right), \label{defA} \\
X &=-2\epsilon_A\epsilon_B  \int_{-\infty}^{\infty} \dif t \! \int_{-\infty}^{t} \dif{t'} \epsilon\left(t \right) \epsilon \left( t' \right) \nn \\
&\qquad \qquad    e^{ i \Omega \left( t +  t' \right)}  W \left(x_A \left(t\right), x_B\left(t'\right) \right) + \mathcal{O}\!\left(\epsilon_0^4\right), \label{defX} \\
C &= \epsilon_A\epsilon_B\int_{-\infty}^{\infty} \dif{t} \! \int_{-\infty}^{\infty} \dif{t'}   \epsilon(t) \epsilon(t') \nn \\
& \qquad \qquad  e^{i  \Omega (t - t')}  W\left(x_A(t') , x_B(t)\right) + \mathcal{O}\!\left(\epsilon_0^4\right),\label{defC} \\
E &=  \abs{X}^2 + A^2 + 2 |C|^2 +\mathcal{O}\!(\epsilon_0^6), \label{defE}
\end{align}
where $W\left(x, x' \right) := \bra{0}\phi\left(x\right) \phi\left(x'\right)\ket{0}$ is the Wightman function. For an explicit derivation of $\rho_{AB}$ to all orders in perturbation theory for the general case when $H_A \neq H_B$ see Appendix \ref{State of the detectors and entanglement measures}.

In four-dimensional Minkowski space the Wightman function for a real massless scalar field is given by \cite{Bogoliubov:1980}
\begin{align}
W_M \left( x,x' \right) &= \frac{1}{4\pi i} \sgn \left(x^0-x'^0\right)\delta\big((x-x')^2\big) \nn \\
&\qquad  -\frac{1}{4 \pi^2 \left( x-x'\right)^2 }, \label{WigtmanFunction}
\end{align}
where $( x-x')^2 = \eta_{\mu \nu}( x^{\mu}-x'^{\mu})( x^{\nu}-x'^{\nu})$.  Following \cite{Bogoliubov:1980,Bogolubov:1990,Scharf:1989} we  treat the Wightman function as a distribution,  regularizing the resulting integrals via the principal value prescription as necessary; see Appendix \ref{The Wightman function} for more details.

We consider the two detectors to be at rest with respect to one another and separated by a distance  $L=|\mathbf{x}_A-\mathbf{x}_B|$. We find the matrix elements of $\rho_{AB}$ to be
 \begin{align}
A_{M} &= \frac{\epsilon_0^2}{4 \pi}\left[e^{-\sigma^2 \Omega^2} -  \sqrt{\pi} \sigma \Omega \erfc \left(\sigma \Omega\right)\right]+ \mathcal{O}\!\left(\epsilon_0^4\right), \label{AinMinkowski} \\
X_{M} &=  \frac{\epsilon_0^2}{4\sqrt{\pi}} \frac{\sigma}{L} i  e^{-\sigma ^2 \Omega ^2 - \frac{L^2}{4\sigma^2}}   \left[1 + \erf\left(i\frac{L}{2\sigma}\right)\right]+ \mathcal{O} \! \left(\epsilon_0^4\right), \label{XinMinkowski} \\
C_M &=  \frac{\epsilon_0^2 }{ 4 \sqrt{\pi}} \frac{\sigma}{L} e^{-\frac{L^2}{4\sigma^2}} \Bigg(    \Im \left[ e^{i  \Omega L} \erf\left( i \frac{L}{2\sigma} + \sigma \Omega\right) \right]  \nn \\
&\qquad \qquad \qquad \qquad \qquad  -    \sin\left( \Omega L\right)  \Bigg) + \mathcal{O}\!\left(\epsilon_0^4\right) \label{Cexp}, \\
E_M &= \abs{X_M}^2 + A_M^2 + 2 C_M^2 + \mathcal{O} \! \left(\epsilon_0^6\right), \label{Eexp}
\end{align}
where $\erf(z)$ is the error function, $\erfc(z) = 1 - \erf(z)$,  and the subscript $M$ denotes that these quantities are calculated in Minkowski space. We {outline} the details of this calculation in Appendix A1 and B.   In the limit of infinite interaction time, i.e., when the two inertial detectors are always on, the resulting transition rate is zero as expected \cite{Birrell:1982}. In the limit of large $L$ the density matrix indeed approaches the direct product $\rho_A\otimes\rho_B$.

\subsection{Information-theoretical properties of the joint state}
\label{Informational properties of the joint state}

Prior to the interaction with the field, the two detectors and the field were initially in a separable state $\ket{\Psi} = \ket{0}_A\ket{0}_B\ket{0}$. After the interaction the detectors become entangled. Since each detector interacts locally with the field, and there is no direct detector-detector interaction, any resulting entanglement between the two detectors must have been redistributed from entanglement already present in the vacuum state of the field. This is the phenomena of entanglement harvesting \cite{Valentini:1991,Reznik:2002fz,MartinMartinez:2012sg, Salton:2014jaa, Pozas-Kerstjens:2015}, and it can be thought of as a kind of entanglement `swapping' from the vacuum state of the field to the joint state of the detectors.

A two-qubit state $\rho$ is entangled if and only if its partially transposed matrix   $\rho^{\Gamma_A}$,
\be
\left(\rho^{\Gamma_A}\right)_{kl,mn}:=\rho_{ml,kn}
\ee
has a negative eigenvalue. This is known as the Peres--Horodecki criterion \cite{Bruss:2007, *Nielsen:2010}. A popular measure of the entanglement between the two detectors, commonly used in the context of entanglement harvesting, is the  negativity
\begin{align}
\mathcal{N}\left(\rho_{AB}\right) := \frac{\left\| \rho^{\Gamma_A} \right\|_1 - 1}{2} =: -\rho_-, \label{defNegativity}
\end{align}
where  $\left\| \cdot \right\|_1$ denotes the trace norm, and $\rho_-$ is the negative eigenvalue in question \cite{Bruss:2007, *Nielsen:2010}.

Another measure of entanglement  that we employ is the concurrence, which plays an important role in entanglement theory and allows us to quantify the amount of entanglement in terms of the  maximally entangled pairs of qubits (ebits) \cite{Bruss:2007, *Nielsen:2010}. For an arbitrary two-qubit state $\rho$ the concurrence is equal to
\begin{align}
\mathcal{C}\left(\rho\right) =\max\left(0,\lambda_1-\lambda_2-\lambda_3-\lambda_4\right), \label{DefConcurrence}
\end{align}
where the $\lambda_i$'s are the square roots of the eigenvalues of the matrix $\rho \tilde{\rho}$, where $\tilde{\rho} = (\sigma_y \otimes \sigma_y ) \rho^* (\sigma_y \otimes \sigma_y )$ with $\sigma_y$ being the Pauli $y$ matrix, ordered such that $\lambda_1\geq \lambda_2\geq \lambda_3\geq \lambda_4$ \cite{Wootters:1997id}. We discuss the relationship between these two measures of entanglement in Appendix \ref{entAB}.

 {From the Peres--Horodecki criterion  it follows that} X-states are entangled if and only if either of the alternatives
\begin{align}
r_{14}^2>r_{22}r_{23}, \qquad r_{23}^2>r_{11}r_{44},
\end{align}
holds.

\begin{figure}
\subfloat[The concurrence $\mathcal{C}\left(\rho_{AB}\right)/\epsilon_0^2$ for detectors in Minkowski space]{%
      \includegraphics[width=0.45\textwidth]{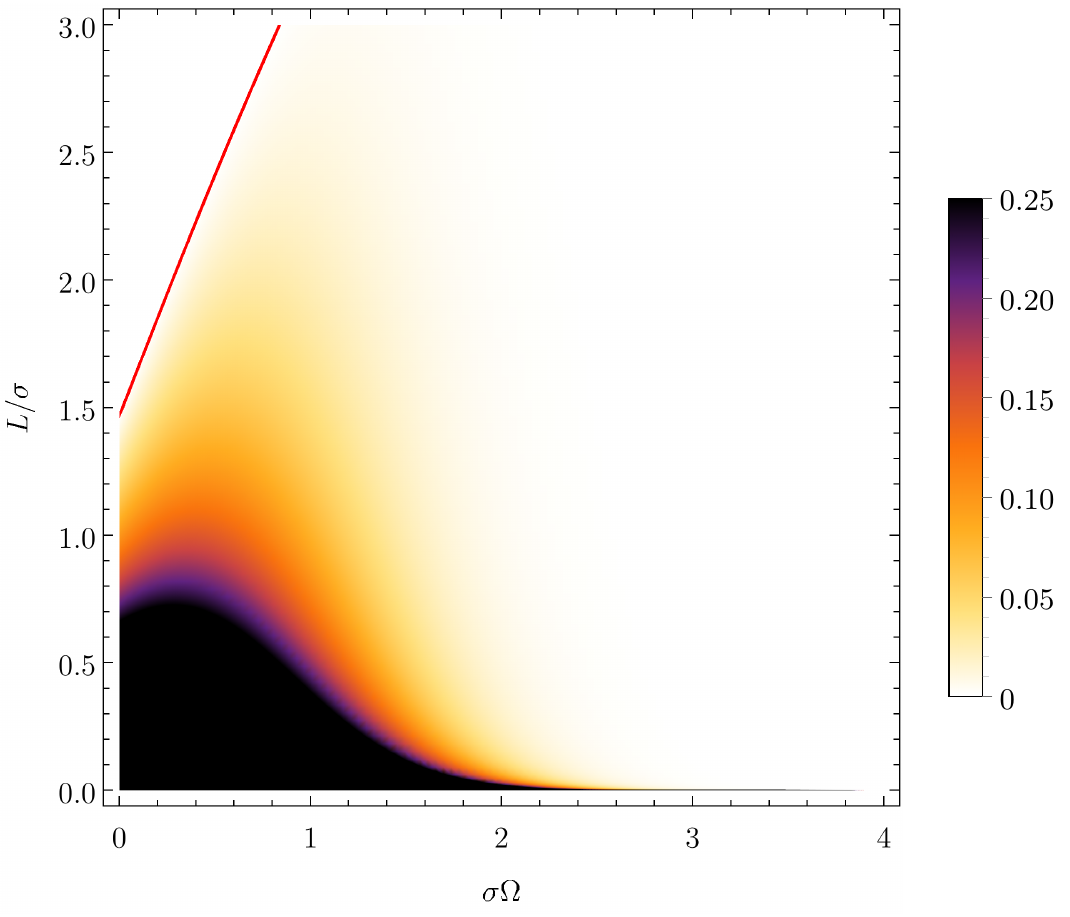} \label{concurr}
    }
    \hfill
    \subfloat[The correlation function $\mathsf{corr}_{AB}/\epsilon_0^2$ for detectors in Minkowski space]{%
      \includegraphics[width=0.45\textwidth]{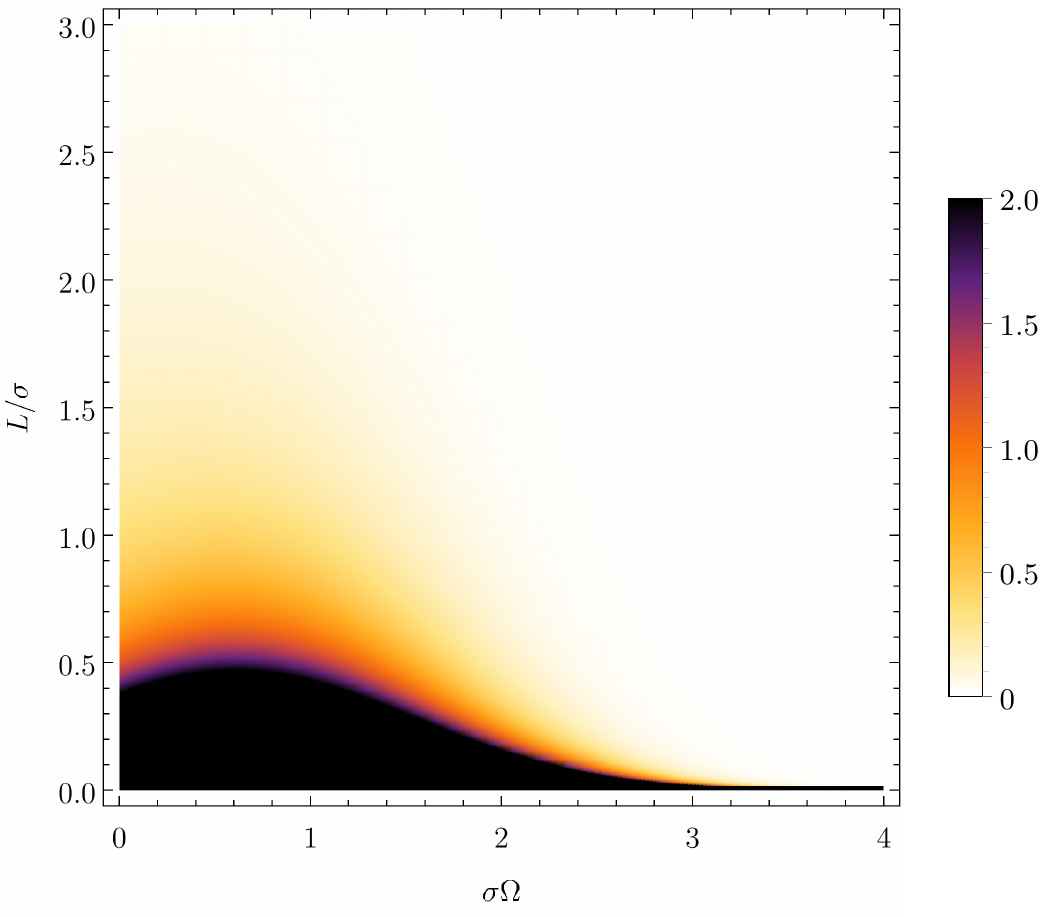} \label{corrplot}
    }
    \caption{The (a) concurrence $\mathcal{C}\left(\rho_{AB}\right)/\epsilon_0^2$ and the (b) correlation function $\mathsf{corr}_{AB}/ \epsilon_0^2$ are plotted for two identical detectors in Minkowski space $\mathcal{M}$, as a function of their separation $L$ and their energy gap $\Omega$ in units of $\sigma$. The red line in (b) corresponds to $\abs{X}_M=A_M$, above which the concurrence vanishes exactly. The solid black regions in the plots indicate the plot has been clipped at (a) $\mathsf{corr}_{AB}/ \epsilon_0^2=2$ and (b) $\mathcal{C}_{\mathcal{M}}\left(\rho_{AB}\right)/\epsilon_0^2 = 0.25$.  }
    \label{fig:CorrelationConcurrence}
\end{figure}

For two identical detectors in the state $\rho_AB$ given in Eq.~\eqref{TwoDetectors}, these conditions are equivalent to
\begin{align}
|X|-A+\co(\epsilon_0^4)>0,\qquad |C|-\sqrt{E} +\mathcal{O}\!\left(\epsilon_0^4\right)  >0,
\end{align}
respectively. However, Eqs.~\eqref{Cexp} and \eqref{Eexp} ensure that the second condition is never satisfied. Hence the entanglement is
\begin{align}
\cc/2=\cn =\max\left(0,|X|-A+\mathcal{O}\!\left(\epsilon_0^4\right) \right), \label{concurrenceMinkowski}
\end{align}
if and only if $r_{14}>r_{22}$.
Using the expressions in Eqs. \eqref{AinMinkowski} and \eqref{XinMinkowski}, we plot the concurrence of the state $\rho_{AB}$ in Fig.~\ref{concurr}.

For a system of two qubits, the  concurrence can be used to  calculate the entanglement of formation $E_F$, defined as the number of the maximally entangled states needed to prepare $\rho_{AB}$ \cite{Bruss:2007, *Nielsen:2010}
\begin{align}
E_F\left(\rho_{AB}\right)&=h\left(\frac{1+\sqrt{1-\mathcal{C}\left(\rho_{AB}\right)^2}}{2}\right) \nn \\
&=\frac{\cc^2}{4\ln 2}\big(1-\ln(\cc^2/4)\big)+ \mathcal{O}\!\left(\epsilon_0^6\right),
\end{align}
where $h(x)=-x\log_2x-(1-x)\log_2(1-x)$. From the above we see that in the standard units of a bipartite entanglement, the so called ebit, the entanglement between the two detectors in a perturbative regime is relatively week: it scales as $\epsilon_0^4$.

{Neither of these quantities is accessible by local measurements of either detector. Instead, we consider two local observers, one in possession of detector $A$ and the other in possession of detector $B$, each measuring the observable Pauli $z$ operator $\sigma_z$, and quantify the correlation in the outcomes of their measurements.} We characterize their results by random variables $r_A$ and $r_B$, respectively, with $r_A, \ r_B \in \{0,1\}$.  The correlation between these variables is given by
\begin{align}
\mathsf{corr}_{AB} =\frac{\mathsf{cov}_{AB}}{\sigma_A\sigma_B} = \frac{E-AB}{\sqrt{A(1-A) B(1-B)}}, \label{corrGeneral}
\end{align}
where $\mathsf{cov}_{AB} :=\braket{r_A r_B}-\braket{r_A}\braket{ r_B}$ is the covariance between $r_A$ and $r_B$ and $\sigma_A^2 = \mathsf{cov}_{AA} $ and $\sigma_B^2 = \mathsf{cov}_{BB}$ are the variances associated with $r_A$ and $r_B$. In  Minkowski space, when the switching function of the two detectors is coincident, the correlation between the measurements performed on the two detectors is given by
\begin{align}
\mathsf{corr}_{AB}^M = \frac{\abs{X_M}^2+2C_M^2}{A_M}+\mathcal{O}\!\left(\epsilon_0^4\right) . \label{corr}
\end{align}



\section{Detectors in quotients of Minkowski space}
\label{Detectors in quotients of Minkowski space}

We now consider an analogous situation in two different quotient spacetimes of Minkowski space with non-trivial spatial topologies. The first is a cylindrical universe $\mathcal{M}_0$
\begin{align}
\mathcal{M}_0 = \mathcal{M}/\Gamma_0,
\end{align}
which is constructed as a quotient of Minkowski space $\mathcal{M}$ with the group $\Gamma_0 = \{ J_0^n \}$ generated by the discrete isometry $J_0: \left( t, x, y ,z \right) \mapsto \left( t, x, y, z + \ell \right)$.

The second spacetime we consider $\mathcal{M}_-$, is again a cylindrical universe in which rotations by $\pi$ in the $xy$-plane have been identified
\begin{align}
\mathcal{M}_- = \mathcal{M}/\Gamma_-,
\end{align}
where the group $\Gamma_- = \{ J_-^n \}$ is generated by the discrete isometry $J_-: \left( t, x, y ,z \right) \mapsto \left( t, -x, -y, z + \ell \right)$.

The compactification scale of these two spacetimes is $\ell$. Both $J_0$ and $J_-$ preserve space and time orientation and act freely and properly \footnote{Let $G$ be a group which acts on a topological space $X$. A group action $G \times X \to X$ is said to act freely if for all $x \in X$, $g x =x$ implies $g$ is the identity. A group action $G \times X \to X \times X$ is said to act properly if the map $(g,x) \mapsto (gx,x)$ is proper, that is inverses of compact sets are compact.}, this ensures that both $\mathcal{M}_0$ and $\mathcal{M}_-$ are space and time orientable Lorentzian manifolds. As neither  $J_0$ and $J_-$ affect the Minkowski line element, both $\mathcal{M}_0$ and $\mathcal{M}_-$ are locally flat spacetimes. The behaviour of particle detectors in both of these spacetimes has been studied in the past, with a focus on applications to Hawking radiation and the Unruh effect \cite{Louko:1998, Langlois:2005}.

The advantage of considering spacetime topologies built from quotients of Minkowski space, is that the Wightman function in the quotient spacetimes can be easily constructed via the method of images from the Wightman function in Minkowski space. If we are given a Green's function $G_{\mathcal{M}}(x,x')$ on a spacetime $\mathcal{M}$, we can construct the corresponding Green's function $G_{\mathcal{M}/J}(x,x')$ on the quotient spacetime $\mathcal{M}/J$ by the image sum \cite{Banach:1979, *Banach:1979a}
\begin{align}
G_{\mathcal{M}/J}(x,x') = \sum_{n=-\infty}^{\infty} \!  \eta^n G_{\mathcal{M}}(x, J^n x'), \label{ImageSum}
\end{align}
where $J x'$ denotes the group action of the group element $J$ on $x'$ and $\eta=\left\{1,-1\right\}$ corresponding to normal scalar fields and twisted fields respectively. From now on we will restrict ourselves to $\eta=1$ for simplicity; the results presented are easily generalizable to twisted fields.

Making use of Eq.~\eqref{ImageSum}, along with the Wightman function in Minkowski space, Eq.~\eqref{WigtmanFunction},  the Wightman function in both $\mathcal{M}_0$ and $\mathcal{M}_-$ can be constructed, and used to evaluate the transition probability $A$ in Eq.~\eqref{defA}, and the matrix elements $X$ and $C$ in Eqs.~\eqref{defX} and\eqref{defC} respectively,  in both spacetimes. We consider two detectors with the world lines
\begin{align}
& x_A(t)=(t, \mathbf{d}_A, z_A),  \qquad    x_B(t)=(t,\mathbf{d}_B, z_B), \label{trajA}
\end{align}
where $\mathbf{d}_A$ and $\mathbf{d}_B$ are two dimensional vectors lying in the $xy$-plane. Matrix elements of $\rho_{AB}$ up to the second order in $\epsilon_0$ in $\mathcal{M}_0$ are given by
\begin{align}
 A_{0} &= A_M +  \frac{\epsilon_0^2}{4 \sqrt{\pi}}  \sigma \sum_{n\neq0} \frac{e^{-\frac{n^2 \ell^2 }{4\sigma ^2}}}{n\ell} \nn \\
 &\  \times \left(   \Im \left[ e^{-i n \ell \Omega }\erf\left(\frac{ n \ell }{2 \sigma }+i \sigma  \Omega \right)\right]   -  \sin (n \ell \Omega)   \right) \label{AinCylindrical} \\
X_0 &=  X_M \nn \\
& \ +  \frac{\epsilon_0^2}{4  \sqrt{\pi}} i e^{-\sigma^2 \Omega^2} \sum_{n\neq0}    \frac{\sigma}{L_n} e^{-\frac{L_n^2}{4\sigma^2}} \left[ 1 + \erf\left( i \frac{L_n}{2 \sigma}\right) \right] \label{XinCylindrical} \\
C_0 &=  C_M +  \frac{\epsilon_0^2 }{ 4 \sqrt{\pi}} \sum_n \frac{\sigma}{L_n} e^{-\frac{L_n^2}{4\sigma^2}} \nn \\
&\ \times \Bigg(    \Im \left[ e^{i  \Omega L_n} \erf\left( i \frac{L_n}{2\sigma} + \sigma \Omega\right) \right]     -    \sin\left( \Omega L_n\right)  \Bigg) \nn \\
 & \ + \mathcal{O}\!\left(\epsilon_0^2\right) \label{CinCylindrical}
\end{align}
where $L_n^2 = L^2 + n^2\ell^2 - 2n \ell \Delta z$ and $\Delta z = z_A - z_B$; while in $\mathcal{M}_-$ these quantities are given by Eqs. \eqref{AinCylindrical}-\eqref{CinCylindrical}, with the substitutions
\begin{align}
\ell &\to \ell_n^2 :=  \ell^2 +4 \frac{d_k^2}{n^2} P(n), \label{elln}\\
L_n &\to \tilde{L}^2_n :=L^2 +n^2 \ell^2 - 2 n \ell \Delta z + 4\mathbf{d}_A \! \cdot \! \mathbf{d}_B P(n) \label{Ltilde},
\end{align}
where $d_k = \| \mathbf{d}_k\|$ with $k \in \{ A,B\}$ and  $P(n) $ is zero or one for even and odd $n$ respectively.

Note that  $A_-$, $X_-$, and $C_-$ for detectors in $\mathcal{M}_-$ depend on the absolute position of the two detectors, as can be seen from the dependence of Eqs. \eqref{elln} and \eqref{Ltilde} on $\mathbf{d}_A$ and $\mathbf{d}_B$. This is a qualitative difference between $\mathcal{M}_-$, and both the cylindrical universe  $\mathcal{M}_0$ and Minkowski space $\mathcal{M}$, stemming from the fact that $\mathcal{M}_-$ is not translationally invariant in the $xy$-plane, which can bee seen from the isometry $J_-$ used to define $\mathcal{M_-}$.  {Consequently, $A_- \neq B_-$, as defined in Eq.~\eqref{TwoDetectors}, which was true for both $\mathcal{M}$ and $\mathcal{M}_0$, and Eq.~\eqref{corrGeneral} must be used to calculate the correlation function.}

From the Eqs. \eqref{AinCylindrical}--\eqref{CinCylindrical} we see for a large compactification scale $\ell$ of the quotient spacetimes, i.e., $\ell\rightarrow \infty$, the contribution to the transition probability and exchange probability from the image sums vanish, resulting in $A_0=A_-=A_M$ and $X_0=X_-=X_M$. Consequently, in this limit the two universes are indistinguishable by measurements of  either the transition probability of a single detector or the resulting entanglement between two detectors.

\begin{figure}
    \subfloat[Transition probability in $\mathcal{M}_0$]{%
      \includegraphics[width=0.45\textwidth]{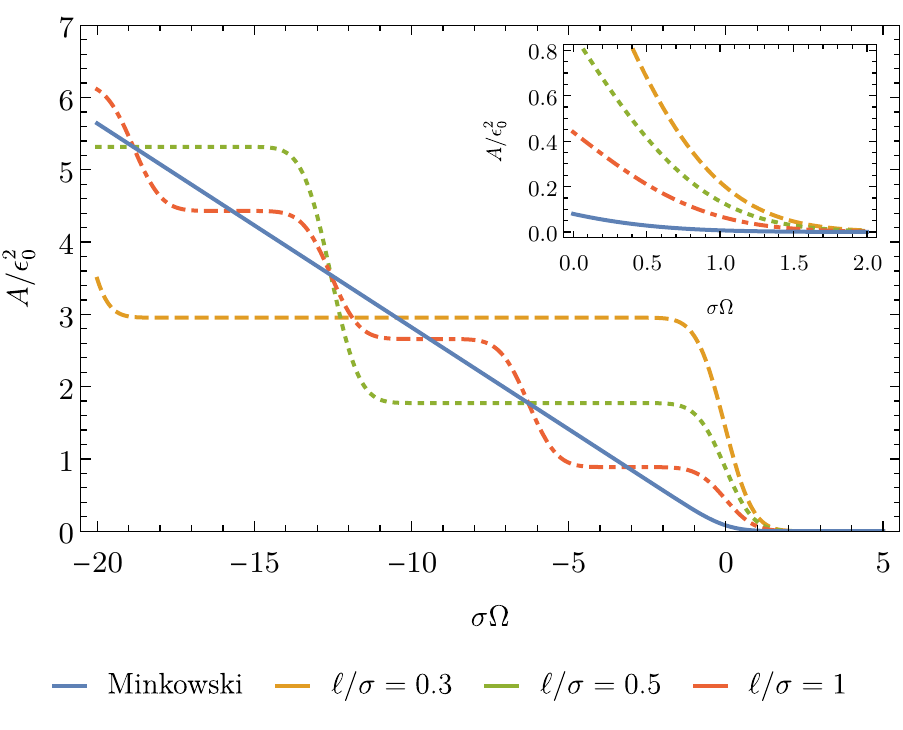}
    }
    \hfill
    \subfloat[Transition probability in $\mathcal{M}_- $ for $d_A/\sigma=1$]{%
      \includegraphics[width=0.45\textwidth]{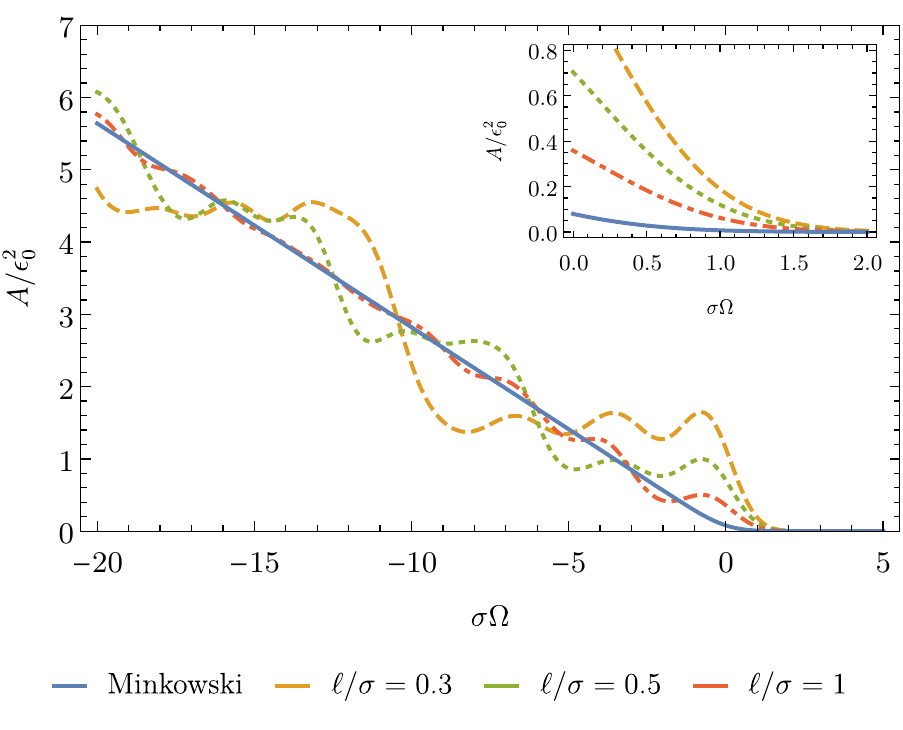}
    }
    \caption{The transition probability of a detector in Minkowski space $\mathcal{M}$ is compared to the transition probability of a detector in both (a) the $\mathcal{M}_0$  spacetime and (b) the $\mathcal{M}_-$ spacetime, by plotting it as a function of the energy gap of the detector, $\Omega$ in units of $\sigma$, for different circumferences $\ell$ of the universe. In both (a) and (b) the dashed blue line is the Minkowski transition probability $A_M$. In the leading order, the probability of the transition $\ket{1} \to \ket{0}$, $E_1-E_2=\Omega>0$,  is equivalent to the excitation probability $ \ket{0} \to \ket{1}$ with $\Omega<0$. {The inset is a magnification of the excitation probability $\Omega>0$.}}
    \label{fig:TransitionProbability}
\end{figure}

\begin{figure}
    \subfloat[Difference in $\mathsf{corr}_{AB}$ between detectors in $\mathcal{M}$ and $\mathcal{M}_0$]{%
      \includegraphics[width=0.45\textwidth]{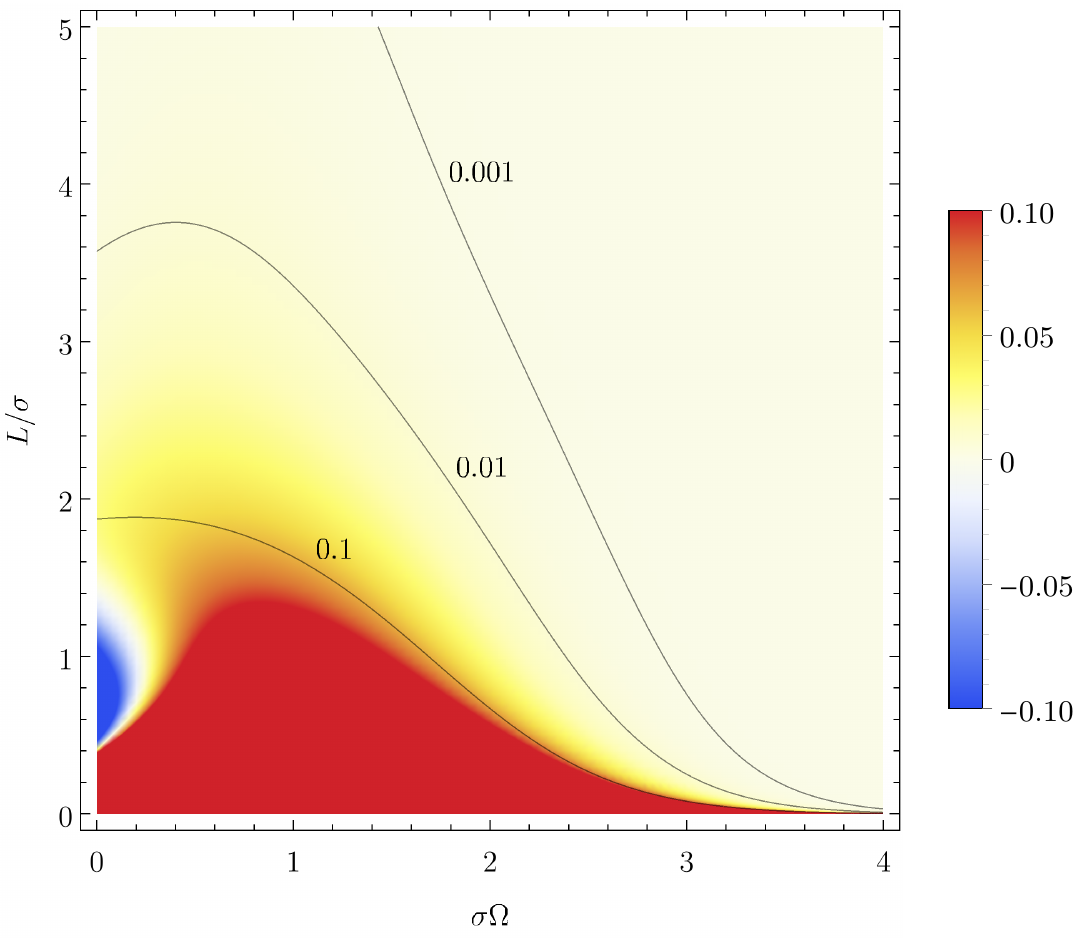}
    }
    \hfill
    \subfloat[Difference in $\mathsf{corr}_{AB}$ between detectors in $\mathcal{M}$ and $\mathcal{M}_-$]{%
      \includegraphics[width=0.45\textwidth]{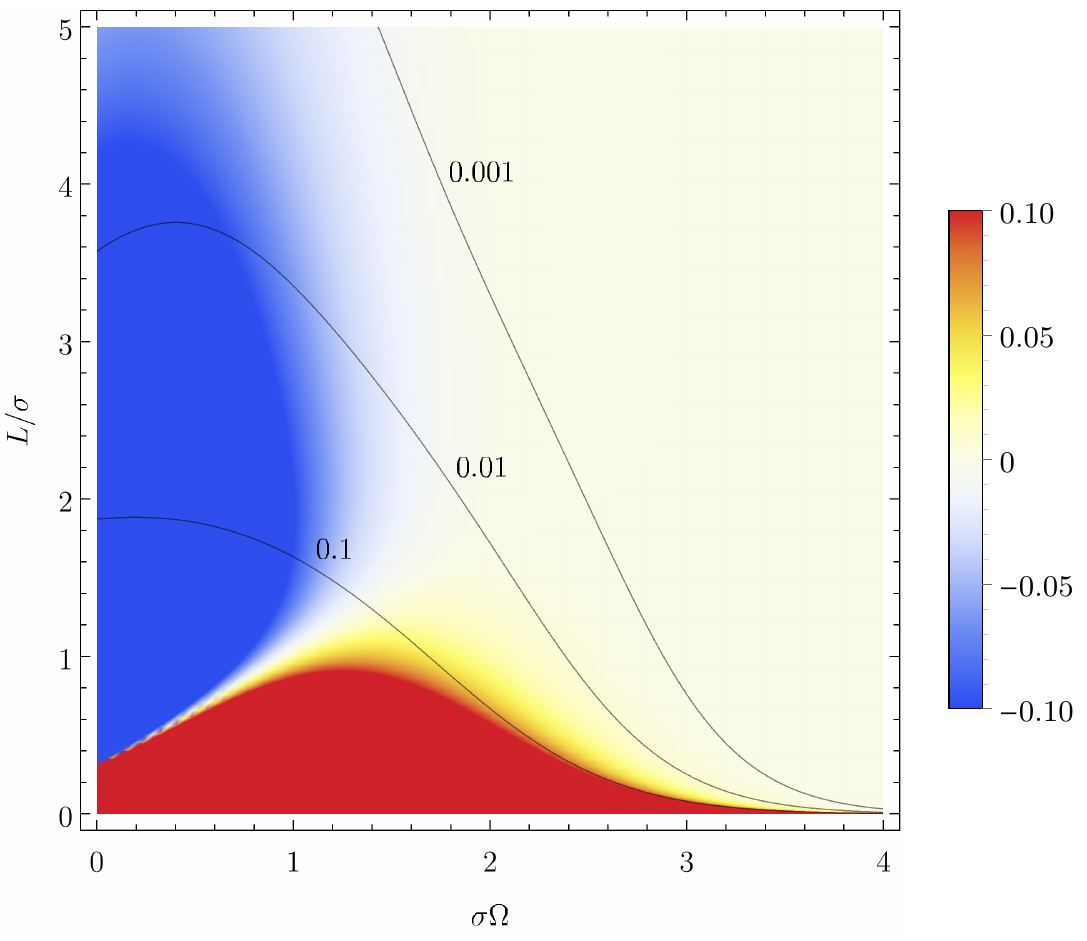}
    }
    \caption{ As a measure of the effect the spatial topology has on the observable correlations in $\rho_{AB}$, we plot the difference between the correlation function for detectors in Minkowski space and (a) the $\mathcal{M}_0$ spacetime: $\mathsf{corr}_{AB}^M/\epsilon_0^2 - \mathsf{corr}_{AB}^{0}/\epsilon_0^2$, and (b) the $\mathcal{M}_-$ spacetime: $\mathsf{corr}_{AB}^M/\epsilon_0^2 - \mathsf{corr}_{AB}^-/\epsilon_0^2$. In addition, we plot the contour lines of the Minkowski correlation function $\mathsf{corr}_{AB}^M/\epsilon_0^2$ to serve as a benchmark. In both plots we have chosen both detectors to lie in the $xy$-plane, so $\Delta z =0$, and choose the length of the identified direction to be $\ell=1$. In plot (b) we have chosen $\mathbf{d}_A$ and  $\mathbf{d}_B$ to be parallel so that $L = d_B-d_A$, and have set $d_A=0.1$. The red region indicates when the correlations between the detectors is greater in Minkowski space than the identified spacetime ($\mathcal{M}_0$ or $\mathcal{M}_-$) and vice versa for the blue region. The solid blue and red regions indicate where the plots have been clipped respectively at $0.10$ and $-0.10$.}
    \label{fig:CorrelationDifference}
\end{figure}

To examine the effect the spatial topology of a spacetime has on the transition probability of a detector, in Fig.~\ref{fig:TransitionProbability} we plot the transition probabilities $A_M$, $A_0$, and $A_-$, in Minkowski space $\mathcal{M}$, the cylindrical spacetime $\mathcal{M}_0$, and the spacetime $\mathcal{M}_-$. The oscillations in the cases of $\mathcal{M}_0$ and $\mathcal{M}_-$ are expected and akin to the appearance of modified quasi-normal modes in spacetimes with closed topologies (see, for instance \cite{Ng2014}).

{The dependence of the transition rate, that is the derivative of the transition probability, on spatial topology for detectors that have been on for an infinite amount of time in both $\mathcal{M}_0$ and $\mathcal{M}_-$ has been studied in \cite{Langlois:2005}. In black hole spacetimes with identical geometry but differing spatial topology, the transition rate of detectors has been studied in  \cite{Smith:2014}. }

In examining the effect the spatial topology has on vacuum entanglement, we plot in Fig.~\ref{fig:CorrelationDifference} the difference in correlations functions between detectors in $\mathcal{M}$ and $\mathcal{M}_0$  and detectors in $\mathcal{M}$ and $\mathcal{M}_-$.

\begin{figure}
\includegraphics[width=0.45 \textwidth]{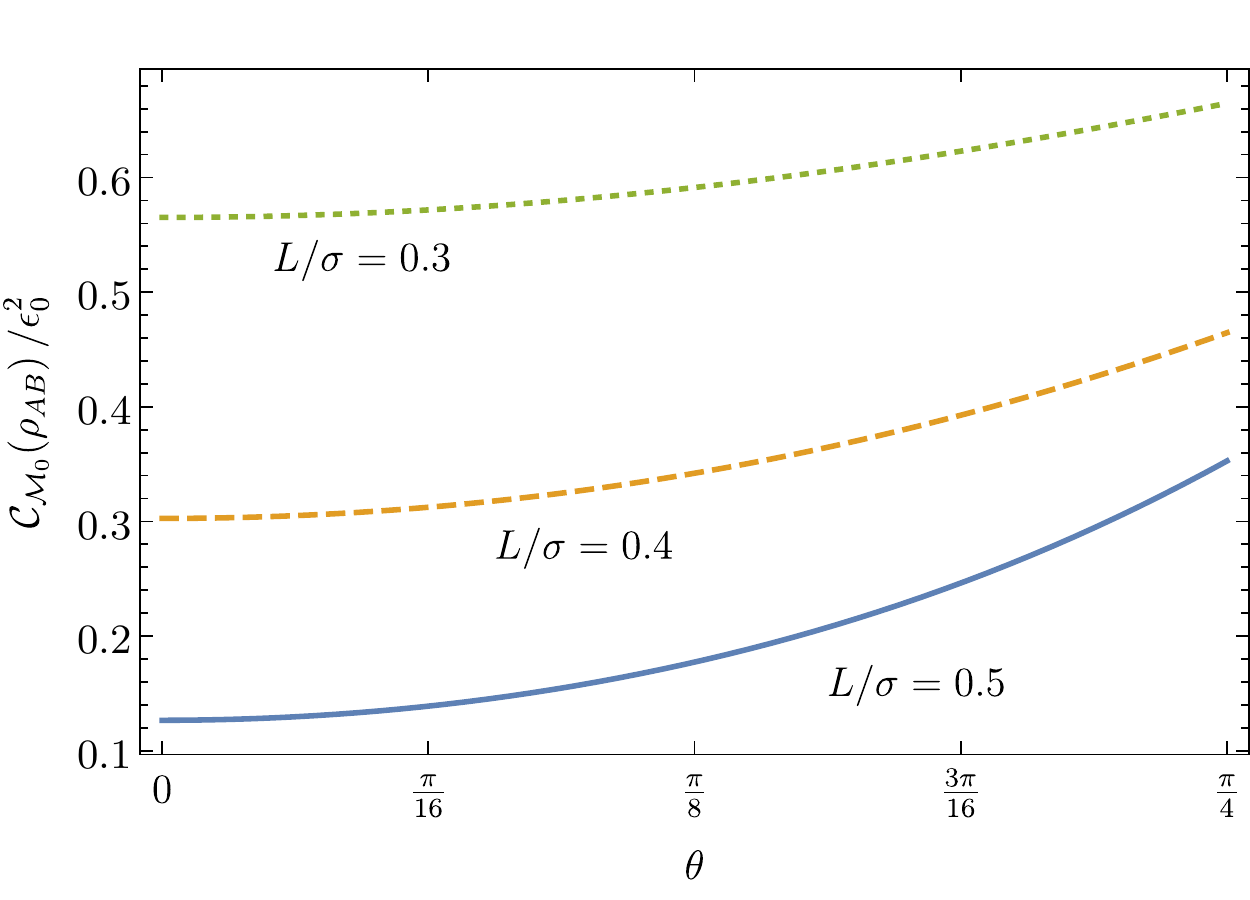}
\caption{ The concurrence of $\rho_{AB}$ for two detectors in the cylindrical universe $\mathcal{M}_0$ with circumference $\ell=1$ is plotted as a function of their orientation $\theta$ with respect to the identified $z$-direction. As $\mathcal{M}_0$ is translational invariant, we can choose our coordinate system such that detector $A$ is at the spatial origin $\mathbf{x}_A=0$ and detector $B$ is located at $\mathbf{x}_B = (L \cos \theta, 0 ,L \sin \theta)$.}
 \label{fig:thetaDependence}
\end{figure}

In both $\mathcal{M}_0$ and $\mathcal{M}_-$ the concurrence of the joint state of the two detectors depends on the orientation of the detectors with respect to the identified direction. Thus, in principle, by measuring vacuum corrections one may infer a preferred direction induced by the spatial topology of the universe. To illustrate this point, in Fig.~\ref{fig:thetaDependence} we plot the dependence of the concurrence of $\rho_{AB}$ on the orientation of the two detectors in the $\mathcal{M}_0$ spacetime with respect to the identified direction.

For the purposes of plotting, in Figs.~\ref{fig:CorrelationConcurrence}-\ref{fig:thetaDependence}, we truncate the image sums at $n = \pm 10$, as the inclusion of more terms does not affect the plots. This is because the contribution from larger $n$ terms in the image sums decease quickly as $n$ grows: $\propto e^{-n^2}/n$, which can be seen from Eqs. \eqref{AinCylindrical}-\eqref{CinCylindrical}.

\section{Discussion and outlook}
\label{Discussion and outlook}


In the spacetimes we analyzed ($\mathcal{M}_0$ and $\mathcal{M}_-$), the effects of global structure show up as small deviations in transition rate of a single particle detector, and entanglement and observable correlations between two detectors from their counterparts in the spatially Euclidean Minkowski spacetime. However, in the limit of zero extrinsic curvature, i.e. when the compactification scale becomes infinite,  these deviations approach zero and the results coincide with that of Minkowski space. We plan to  disentangle the role of extrinsic curvature  from that of entirely topological effects in future work.

The Minkowskian result is recovered in the limit $\ell/\sigma\rightarrow\infty$. Since in cosmological scenarios we expect the topological scale to be at the order of the Hubble scale, $\ell\sim 1/H$ and the effective experimental run-time is at the order of years at  best, we expect $\ell/\sigma\succeq 10^{10}$, making the effects unobservable.  By the same token we expect that entanglement between different degrees of freedom during early  Universe, where the relevant dimensionless ratios were of order of 1 or higher, {to be} significantly impacted by the emerging global structures. We plan to investigate these effects in future work.

{On the other hand, although the discussion throughout this paper has focused on the spatial topology of the entire Universe, the tools used and results presented apply equally well to fields and detectors in cavities with appropriate boundary conditions. Specifically, the results given regarding the cylindrical universe $\mathcal{M}_0$ are equivalent to detectors and the field in a cavity with periodic boundary conditions. Further, the blue regions in Fig.~\ref{fig:CorrelationDifference} suggest that entanglement harvesting from a quantum field may be increased by constructing cavities with appropriate boundary conditions. {In fact,  there is already substantial evidence that this is the case: Entanglement harvesting in cavities has already been analyzed non-perturbatively in \cite{Non-pertur} and further in \cite{Martin-MartinezSUS}, where it was shown that a combination of harvesting in cavity setups complemented with communication yields a sustainable source of quantum entanglement. This particular amplification of harvesting in cavity setups is otherwise impossible in free space.}}

A different group of questions deals with  entanglement and correlations between the detectors in relative motion \cite{Peres:2004}, as well as with the effects of delay between switching the detectors. Finally, it is interesting to investigate the build-up of the correlations between the detectors in time.

The density matrix of Eq.~\eqref{rhoAB} has the same form at all orders of perturbation theory. Nevertheless, it will be instructive to obtain non-perturbative results that are based on non-perturbative methods following the several different formalisms developed for  harmonic oscillator detectors  \cite{linhu:06, Non-pertur, Fuenetesnonpert}, as well is to compare different types of detectors.

\medskip

\begin{acknowledgments}
The authors would like to thank Nicolas C. Menicucci and Robert B. Mann for useful discussions.
\end{acknowledgments}

\appendix


\section{State of the detectors and entanglement measures}
\label{State of the detectors and entanglement measures}

\subsection{Calculation of $\rho_{AB}$}

Here we provide the details of the perturbative calculations that lead to the reduced density matrix  in Eq. \eqref{TwoDetectors}. We present the derivation in Minkowski space, but it translates verbatim to any stationary spacetime.

Here we work in the Scr\"{o}dinger picture as it will be more convenient than the interaction picture to extract the structure of the reduced density matrix $\rho_{AB}$ to all orders in perturbation theory. The Hamiltonian describing two detectors interacting with the scalar field $\phi$ is given by
\begin{align}
H=H_A + H_B+H_\phi+H_{\rm int},
\end{align}
where $H_A=\tfrac{1}{2}\Omega_A \sigma_{z}$ and $H_B=\tfrac{1}{2}\Omega_B \sigma_{z}$ are the free Hamiltonians of detectors $A$ and $B$ with energy gaps $\Omega_A$ and $\Omega_B$, $H_\phi$ is the free Hamiltonian of the scalar field, and $H_{\rm int}$ is the interaction Hamiltonian in the Sch\"{o}rdinger picture
\begin{align}
H_{\rm int}(t)=\epsilon_A(t)\sigma_x\phi_S\big(x_A(t)\big)+\epsilon_B(t)\sigma_x\phi_S\big(x_B(t)\big). \label{interaction2detectors}
\end{align}
where the subscript $S$ reminds us that we are working in the Scr\"{o}dinger picture. We choose the switching functions $\epsilon_A(t)$ and $\epsilon_B(t)$ to be proportional to our perturbation parameter $\epsilon_0\ll1$; in the analysis presented we took $\epsilon_A(t)=\epsilon_B(t) = e^{-t^2/2\sigma^2}$, but in general other switching functions may be studied. Since both detectors are inertial and at rest with respect to one another, we have equated the proper time of each detector with the coordinate time $t$.

Initially ($t_\mathrm{in}\ra -\infty$) the detectors are in their ground state and the field in the vacuum state
\begin{align}
\ket{\Psi}=\ket{0}_A\ket{0}_B\ket{0}. \label{incon}
\end{align}
\begin{widetext}
We  expand the state $\ket{\psi(t)} = \exp(-iHt) \ket{\Psi}$ of the combined system in the eigenstates of the unperturbed Hamiltonian at time $t$ as
\begin{align}
\ket{\psi\left(t\right)}=&  \SumInt_{m_A m_B\, \mu} \Big( c_{m_A m_B\, \mu}(t)e^{-i(E_{m_A}+E_{m_B})t}  U(t) \ket{m_A,m_B,\mu} \Big),
\end{align}
where $m_i\in \left\{0,1\right\}$ labels the energy levels of the detectors with energies $E_{1_A} - E_{0_A} =\Omega_A$ and $E_{1_B} - E_{0_B} =\Omega_B$, the index $\mu$ represents a decomposition over the basis states of the scalar field, and $U(t)=\1_A \otimes \1_B\otimes U_\phi(t)$ where $U_\phi(t) = \exp(-i H_\phi t)$. The coefficients $c_{m_A m_B\, \alpha}(t)$ satisfy the Schr\"{o}dinger equation
\begin{align}
i\dot{c}_{m_A m_B\,\mu} \left(t\right) &= \SumInt_{n_A,\nu} \Big( e^{i(E_{m_A}-E_{n_A})t} \epsilon_A(t)(\sigma_x)_{m_An_A}     \Braket{ \mu|\phi\left[x_A(t)\right]|\nu}c_{n_A m_B\, \nu}\left(t\right) \Big)\nn \\
& \qquad +\SumInt_{n_B,\nu} \Big(e^{i(E_{m_B}-E_{n_B})t}\epsilon_B(t)(\sigma_x)_{m_B n_B}    \Braket{ \mu|\phi\left[x_B(t)\right] |\nu}c_{m_An_B\,\nu}\left(t\right) \Big), \label{expansion}
\end{align}
where $\sigma_x$ is the Pauli $x$ matrix. We represent the coefficients $c_{m_A m_B \mu}$ in a more explicit form as
\be
c_{m_A m_B \mu}=(c_{mn\, \mathrm{v}},c_{mn\, p},c_{mn\,p_1 p_2},\ldots),
\ee
i.e., spell out explicitly the vacuum, one-particle, two-particle, etc. components. To simplify the notation we  label the terms $c_{01\,p}$ as $\vec{b}(p)$ and $c_{11\,p_1p_2}$ as $\vec{x}(p_1,p_2)$, and suppress the arguments.  Contributions of these elements to the inner product that is performed with the appropriate measure will be denoted in the usual vector form, such as $|\vec{x}|^2$ or $\vec{b}\cdot\vec{a}^*$.

The solution to Eq.~\eqref{expansion},  subject to the initial condition given in Eq.~ Eq.~\eqref{incon}, is given schematically as
\begin{align}
& c_{00\alpha} =\big(1+\epsilon_0^2g^{(2)}_\mv e^{i\gamma_2}+
\epsilon_0^4g^{(4)}_\mathrm{v}e^{i\gamma_4},0,\epsilon_0^2\vec{g}^{(2)}+\co(\epsilon_0^4),0,\co(\epsilon_0^4)\big) +\co(\epsilon_0^6)\\
& c_{01\alpha}=\big(0, \epsilon_0 \vec{b}^{(1)}+\epsilon_0^3 \vec{b}^{(3)},0,\co(\epsilon_0^3)\big) +\co(\epsilon_0^5), \\
& c_{10\alpha}=\big(0, \epsilon_0 \vec{a}^{(1)}+\epsilon_0^3 \vec{a}^{(3)},0,\co(\epsilon_0^3)\big) +\co(\epsilon_0^5), \\
& c_{11\alpha}=\big(\epsilon_0^2x^{(2)}_\mathrm{v}e^{i\chi_2}+\epsilon_0^4x^{(4)}_\mathrm{v}e^{i\chi_4},0,\epsilon_0^2\vec{x}^{(2)}\big)+\co(\epsilon_0^4).
\end{align}

As a result, the non-zero matrix elements of the reduced density matrix $\rho_{AB}$ in Eq.~ \eqref{rhoAB} describing the joint state of the two detectors, up to the fourth order in $\epsilon_0$ are given by
\begin{align}
&\rho_{00,00}=: r_{11}=1+2\epsilon_0^2 g^{(2)}_\mv\cos\gamma+\epsilon_0^4\left[\big(g^{(2)}_\mv\big)^2+2g^{(4)}_\mv\cos\gamma_4+\big|\vec{g}^{(2)}\big|^2\right], \\
&\rho_{01,01}=: r_{22}=
\epsilon_0^2\big|\vec{a}^{(1)}\big|^2+\epsilon_0^4\big(\vec{a}^{(1)}\cdot\vec{a}^{(1)*}+\vec{a}^{(1)*}\cdot\vec{a}^{(1)}\big), \\
&\rho_{10,10}=: r_{33}=
\epsilon_0^2\big|\vec{b}^{(1)}\big|^2+\epsilon_0^4\big(\vec{b}^{(1)}\cdot\vec{b}^{(1)*}+\vec{b}^{(1)*}\cdot\vec{b}^{(1)}\big), \\
&\rho_{11,11}=:r_{44}=\epsilon_0^4\left[\big(x^{(2)}_\mv\big)^2+\big|\vec{x}^{(2)}\big|^2\right], \\
&\rho_{01,10}=: r_{23}e^{-i\zeta}=\epsilon_0^2\vec{b}^{(1)}\cdot\vec{a}^{(1)*}+\epsilon_0^4\big(\vec{b}^{(1)}\cdot\vec{a}^{(3)*}+\vec{a}^{(1)}\cdot\vec{b}^{(3)*}\big),\\
&\rho_{00,11}=: r_{14}e^{-i\xi}=\epsilon_0^2x^{(2)}_\mathrm{v}e^{-i\chi_2}+
\epsilon_0^4\left(g^{(2)}_\mv x^{(2)}_\mathrm{v}e^{i(\gamma_2-\chi_2)}+\vec{g}^{(2)}\cdot\vec{x}^{(2)*}\right).
\end{align}
By construction $\rho_{AB}$ is normalized.

To calculate the transition probability, correlations, and measures of entanglement in the leading order, we need the coefficients $c_{m_A m_B \mu}$ up to order $\epsilon_0^2$:
\begin{align}
& c_{0 1\, p}^{(1)}  = -i \int_{-\infty}^{\infty} dt \,   e^{i  \Omega_B t }  \epsilon_B(t) \Braket{ p |\phi\big(x_B(t)\big) | 0} \label{c01}, \\
& c_{1 0\, p}^{(1)}  = -i \int_{-\infty}^{\infty} dt \,  e^{i \Omega_A t }  \epsilon_A(t) \Braket{ p|\phi\big(x_A(t)\big) | 0},  \\
&c_{0 0\, \nu}^{(2)} =  - \int_{-\infty}^{\infty} dt \int_{-\infty}^{t} dt' \, \Bigg[    \epsilon_A(t) \epsilon_A(t')    e^{-i \Omega_A (t-t') }   \Braket{ \nu|\phi\big(x_A(t)\big) \phi\big(x_A(t')\big) | 0}   \nn \\
& \qquad \qquad \qquad \qquad \qquad \qquad \qquad \qquad +       \epsilon_B(t) \epsilon_B(t')   e^{-i  \Omega_B (t-t') }  \Braket{ \nu|\phi\big(x_B(t)\big) \phi\big(x_B(t')\big) | 0}\Bigg] , \\
&c_{1 1\, \nu}^{(2)}=  - \int_{-\infty}^{\infty} dt \int_{-\infty}^{t} dt' \, \Bigg[     \epsilon_A(t)\epsilon_B(t')  e^{i \Omega_A t +  i  \Omega_B t' }   \Braket{ \nu|\phi\big(x_A(t)\big)  \phi\big(x_B(t')\big) | 0}  \nn \\
& \qquad \qquad \qquad \qquad \qquad \qquad \qquad \qquad+      \epsilon_B(t) \epsilon_A(t') e^{i  \Omega_B t + i \Omega_A t'  } \Braket{ \nu|\phi\big(x_B(t)\big)  \phi\big(x_A(t')\big) | 0}\Bigg] \label{c11},
\end{align}
where in the second order coefficients the index $\nu$ stands either for vacuum (0) or a two-particle state ($p_1p_2$). $\rho_{10,10}$, $\rho_{01,01}$, and $\rho_{00,11}$

{The matrix elements of $\rho_{AB}$ are obtained from Eqs. \eqref{c01}-\eqref{c11} with the help of Eq.~ \eqref{MatrixElements}. For identical detectors the matrix elements $\rho_{10,10} = \rho_{01,01}$, $\rho_{01,10}$, and $\rho_{00,11}$ are obtained straight forwardly from Eqs.  \eqref{c01}--\eqref{c11}, and result in the quantities $A$, $X$, and $C$ respectively, and are given in Eqs. \eqref{defA}--\eqref{defC}.}

{As for the matrix element $\rho_{11,11}$, applying Eq.~ \eqref{MatrixElements} results in
\begin{align}
\rho_{11,11} 
&= \int_{-\infty}^{\infty} dt \int_{-\infty}^{t}dt' \int_{-\infty}^{\infty} dt'' \int_{-\infty}^{t''} dt'''    \epsilon(t)\epsilon(t') \epsilon(t'') \epsilon (t''')  e^{i \Omega( t+t'-t''-t''')}  \Bigg[  \nn \\
& \quad  \Braket{ 0 | \! \phi\big(x_B(t''')\big) \phi\big(x_A(t'')\big)\phi\big(x_A(t)\big)  \phi\big(x_B(t')\big) \!| 0}
+  \Braket{ 0 |\!  \phi\big(x_A(t''')\big) \phi\big(x_B(t'')\big)\phi\big(x_A(t)\big)  \phi\big(x_B(t')\big) \!| 0} \nn \\
& \quad  +   \Braket{ 0 | \!  \phi\big(x_B(t''')\big) \phi\big(x_A(t'')\big)\phi\big(x_B(t)\big)  \phi\big(x_A(t')\big)\! | 0}
+  \Braket{ 0 | \! \phi\big(x_A(t''')\big) \phi\big(x_B(t'')\big)\phi\big(x_B(t)\big)  \phi\big(x_A(t')\big) \!| 0} \Bigg]. \label{E1}
\end{align}
The four 4-point functions appearing in Eq.~ \eqref{E1} can be expanded in terms of products of Wightman functions using the commutation properties of the field, yielding
\begin{align}
\rho_{11,11} &= \int_{-\infty}^{\infty} dt \int_{-\infty}^{t}dt' \int_{-\infty}^{\infty} dt'' \int_{-\infty}^{t''} dt'''
  \epsilon(t)\epsilon(t') \epsilon(t'') \epsilon (t''')  e^{i \Omega( t+t'-t''-t''')}   \Bigg( 4 W \big( x_A(t''') , x_B(t'') \big)  W\big(x_A(t),x_B(t') \big)   \nn \\
& \quad + \Big[W \big( x_A(t''),x_A(t)\big) W \big( x_B(t'''),x_B(t')\big) +   W\big(x_A(t'''),x_A(t) \big) W\big(x_B(t''), x_B(t') \big) \nn \\
& \qquad \qquad + W\big(x_A(t''),x_A(t') \big) W\big(x_B(t'''),x_B(t) \big)  +  W \big( x_A(t''') , x_A(t')\big) W \big( x_B(t'') , x_B(t)\big)\Big]\nn \\
& \quad  + 2  \Big[  W\big(x_B(t'''),x_A(t) \big) W\big(x_A(t''),x_B(t') \big) + W \big( x_B(t''),x_A(t)\big)W \big( x_A(t'''), x_B(t') \big)  \Big] \Bigg), \label{E2}
\end{align}}
\end{widetext}
{where we have exploited the fact that the detectors are at rest with respect to one another, which results in $W\big(x_A(t), x_B(t') \big) = W\big(x_B(t), x_A(t') \big)$.}

{The first term appearing in Eq.~ \eqref{E2} is immediately identified as $\abs{X}^2$, where $X$ is given in Eq.~ \eqref{defX}. By changing the integration variables to
\begin{align}
u &= t'''-t,   &v &= t''-t', \nn \\
\bar{u} &= t'''+t, &\bar{v} &= t''+t', \label{coordinates1}
\end{align}
and then to
\begin{align}
a &= \bar{v} - \bar{u} & b &= \bar{v} + \bar{u},
\end{align}
the second and third terms in Eq.~ \eqref{E2} result in $AB$ and $2C^2$ respectively, where $A$ and $B$ are defined in Eq.~ \eqref{defA} and $C$ in Eq.~ \eqref{defC}. Thus we find
\begin{align}
\rho_{11,11} = \abs{X}^2 + AB + 2 C^2.
\end{align}}

{The matrix element $\rho_{00,00}$ is obtained either directly from its definition or by exploiting the normalization condition $\tr ( \rho_{AB}) =1$. }

\subsection{Quantifying the entanglement in $\rho_{AB}$}\label{entAB}

We give here negativity and concurrence for the density matrix of \eqref{rhoAB}. In the case $r_{14}^2>r_{22}r_{23}$
\begin{align}
\cn=-\tfrac{1}{2}(r_{22}+r_{33}-\sqrt{(r_{22}-r_{33})^2+4r_14^2}),
\end{align}
and
\begin{align}
\cc=2(r_{14}-\sqrt{r_{22}r_{23}}),
\end{align}
which simplifies further if  $r_{22}=r_{23}$  to
\begin{align}
\cn=\cc/2=r_{14}-r_{22}.
\end{align}

If $r_{23}^2>r_{11}r_{44}$ then
\begin{align}
\cn=-(r_{11}+r_{44}-\sqrt{(r_{11}+r_{44})^2+4r_{23}}),
\end{align}
and
\begin{align}
\cc=2\left(r_{23}-\sqrt{r_{11}r_{44}}\right).
\end{align}

In a general two-qubit system concurrence and negativity are related by \cite{vadm:01}
\begin{align}
\cc\geqslant 2\cn\geqslant\sqrt{(1-\cc)^2+\cc^2}+(1-\cc).
\end{align}
The negativity is equal to the concurrence if the eigenvector of the partially transposed state $\rho_{AB}^{\Gamma_A}$ corresponding to its negative eigenvalue is one of the Bell states (up to local unitary transformations).  Indeed, for the identical detectors ($r_{22}=r_{23}$) when $r_{14}>r_{22}$ the two quantities coincide and the eigenvector in question becomes $\tfrac{1}{\sqrt{2}}(0,-1,1,0)^T$.

\section{The Wightman function}
\label{The Wightman function}

To ease comparison between different sources, we first spell out how our convention for the Wightman function $W:=\6 0|\phi(x)\phi(y)|0\9$ relates to the definitions  in other references we use. In particular,
\begin{align}
W(x,x')=G^+(x,x')=-iD^-(x,x')
\end{align}
where $G^+$ is introduced in \cite{Birrell:1982} (and is called $D^+$ in the massless case), $D^-$ is introduced in \cite{Bogoliubov:1980}, and is called $D^+$ in \cite{Scharf:1989}.

Wightman functions are well-defined distributions, i.e. they can be represented as a distributional limit of regular analytic functions \cite{Bogoliubov:1980}. However, representation in terms of functions that are also covariant requires a regularization procedure, e.g., the Pauli-Villars regularization. A simple popular representation uses the ``$-i\epsilon$" prescription \cite{Birrell:1982}
\begin{align}
W_\epsilon(x-x')=-\frac{1}{4\pi^2}\frac{1}{(t-t'-i\epsilon)^2-|\mathbf{x}-\mathbf{x}' |^2}. \label{iepsilonW}
\end{align}
Eq.~\eqref{iepsilonW} is not manifestly covariant, and requires additional manipulations to obtain, for example, the Unruh effect \cite{Schlicht:2003iy}. On the other hand, a straightforward calculation demonstrates that the distributional form of the Wightman function as given by Eq.~\eqref{WigtmanFunction} is free from these complications.

For convenience, we summarize here several properties of distributions that we employed in obtaining Eqs.~\eqref{AinMinkowski} and \eqref{XinMinkowski} \cite{Bogolubov:1990}. Recall that the definition of a distribution $G$ acting on a test function $f$ is given by
\begin{align}
\Braket{G, f } :=  \int_{-\infty}^{\infty} g(y) f(y) \, dy, \label{defDistribution}
\end{align}
where the function $g(y)$ defines the distribution $G$. The derivative of a distribution is obtained from the above definition by integrating by parts to give
\begin{align}
\Braket{G', f } = -\Braket{G, f' }. \label{distributionalDerivative}
\end{align}

The distribution $1/x$ acting on  a test function $f(x)$ is defined as
\begin{align}
\Braket{\frac{1}{x},f(x)}=\mathrm{PV}\int_{-\infty}^\infty\!dx\,\frac{f(x)}{x},
\end{align}
where $\PV$ denotes that the principle value of the integral should be taken. All the subsequent inverse power distributions $1/x^n$ are defined as distributional derivatives of $1/x$, hence
\begin{align}
\Braket{\frac{1}{x^2},f(x)}&=\Braket{\frac{1}{x},f'(x)} \nn \\
&=\int_0^\infty \dif{x} \frac{f(x)-f(-x)-2f(0)}{x^2}. \label{Idenity1}
\end{align}
Eq.~ \eqref{Idenity1} is used in arriving at the expression for $X_{\mathcal{M}}$ given in \eqref{XinMinkowski}.

Particular care is required for the evaluation of integrals involving the delta-function term of the Wightman function when $y:=x-x'=0$. We consider the distribution $\delta(y^2)$ as a limit of
\begin{align}
\delta\left(y^2\right)&=\lim_{r\rightarrow 0}\delta \left(y^2-r^2\right) \nn \\
&=\lim_{r\rightarrow 0}\frac{1}{2r} \Big(\delta(y-r)+\delta(y+r)\Big).
\end{align}
This prescription results in the action of the distribution $\sgn(y)\delta(y^2)$ on a test function $f(y)$ to be
\begin{align}
 \PV \int_{-\infty}^{\infty} \dif{y} \sgn\left(y\right) \delta\left(y^2\right)  f\left(y\right)
&= f'\left(0\right). \label{Idenity2}
\end{align}
Eq.~ \eqref{Idenity2} is used in arriving at the expression for $A_{\mathcal{M}}$ given in \eqref{AinMinkowski}.

\bibliography{STSE.bib}

\end{document}